\documentclass[prb,twocolumn,superscriptaddress,showpacs]{revtex4}
\usepackage{bm}
\usepackage{amsmath}
\usepackage{amssymb}
\usepackage[dvips]{graphicx}
\begin{document}

\title{\bf Luttinger Liquid and Polaronic Effects in Electron Transport through a Molecular Transistor}
\author{G. A. Skorobagatko}
\email{gleb_skor@mail.ru}
\affiliation{B. Verkin Institute for Low Temperature Physics and Engineering of
the National Academy of Sciences of Ukraine, 47 Lenin Avenue, Kharkov 61103, Ukraine}
\author{I. V. Krive}
\affiliation{B. Verkin Institute for Low Temperature Physics and Engineering of
the National Academy of Sciences of Ukraine, 47 Lenin Avenue, Kharkov 61103, Ukraine}

\date{\today}

\begin{abstract}
Electron transport through a single-level quantum dot weakly
coupled to Luttinger liquid leads is considered in the master
equation approach. It is shown that for a weak or moderately
strong interaction the differential conductance demonstrates
resonant-like behavior as a function of bias and gate voltages.
The inelastic channels associated with vibron-assisted electron
tunnelling can even dominate electron transport for a certain
region of interaction strength. In the limit of strong interaction
resonant behavior disappears and the differential conductance
scales as a power low on temperature (linear regime) or on bias
voltage (nonlinear regime).
\end{abstract}


\maketitle

\section{\bf Introduction}

Last years electron transport in molecular transistors became a
hot topic of experimental and theoretical investigations in
nanoelectronics (see e.g. \cite{1,2}). From experimental point of
view it is a real challenge to place a single molecule in a gap
between electric leads and to repeatedly measure electric current
as a function of bias and gate voltages. Being in a gap the
molecule may form chemical bonds with one of metallic electrodes
and then a considerable charge transfer from the electrode to the
molecule takes place. In this case one can consider the trapped
molecule as a part of metallic electrode and the corresponding
device does not function as a single electron transistor (SET).
Much more interesting situation is the case when the trapped
molecule is more or less isolated from the leads and preserves its
electronic structure. In a stable state at zero gate voltage the
molecule is electrically neutral and the chemical potential of the
leads lies inside the gap between HOMO (highest occupied molecular
orbital) and LUMO (lowest unoccupied molecular orbital) states.
This structure demonstrates Coulomb blockade phenomenon \cite{3,4}
and Coulomb blockade oscillations of conductance as a function of
gate voltage (see review papers in \cite{5} and references
therein). In other words a molecule trapped in a potential well
between the leads behaves as a quantum dot and the corresponding
device exhibits the properties of  SET. The new features in a
charge transport through molecular transistors as compared to the
well-studied semiconducting SET appear due to "movable" character
of the molecule trapped in potential well (the middle electrode of
the molecular transistor). Two qualitatively new effects were
predicted for molecular transistors: (i) vibron-assisted electron
tunnelling (see e.g. \cite{6,7}) and, (ii) electron shuttling
\cite{8} (see also review \cite{9} ).

Vibron(phonon)-assisted electron tunnelling is induced by the
interaction of charge density on the dot with local phonon modes
(vibrons) which describe low-energy excitations of the molecule in
a potential well. This interaction leads to satellite peaks (side
bands) and unusual temperature dependence of peak conductance in
resonant electron tunnelling \cite{10}. For strong electron-vibron
interaction the exponential narrowing of level width and as a
result strong suppression of electron transport (polaronic
blockade) was predicted \cite{10,11}. The effect of electron
shuttling appears at finite bias voltages when additionally to
electron-vibron interaction one takes into account coordinate
dependence of electron tunnelling amplitude \cite{8,9}.

Recent years carbon nanotubes are considered as the most promising
candidates for basic element of future nanoelectronics. Both
$C_{60}$-based and carbon nanotube-based molecular transistors
were already realized in experiment \cite{12,13}. The low-energy
features of I-V characteristics measured in experiment with
$C_{60}$-based molecular transistor \cite{12} can be theoretically
explained by the effects of vibron-assisted tunnelling \cite{7}.

It is well known that in single-wall carbon nanotubes (SWNT)
electron-electron interaction is strong and the electron transport
in SWNT quantum wires is described by Luttinger liquid theory.
Resonant electron tunnelling  through a quantum dot weakly coupled
to Luttinger liquid leads for the first time was studied in
Ref.\cite{14} were a new temperature scaling of maximum
conductance was predicted: $G(T)\propto T^{1/g-2}$ with
interaction dependent exponent (g is the Luttinger liquid
correlation parameter).

In this paper we generalize the results of Refs.\cite{10,14} to
the case when a quantum dot with vibrational degrees of freedom is
coupled to Luttinger liquid quantum wires. The experimental
realization of our model system could be, for instance,
$C_{60}$-based molecular transistors with SWNT quantum wires.

In our model electron-electron and electron-phonon interactions
can be of arbitrary strength while electron tunnelling amplitudes
are assumed to be small (that is the vibrating quantum dot is
weakly coupled to quantum wires). We will use master equation
approach to evaluate the average current and noise power. For
noninteracting electrons this approximation is valid for
temperatures $T\gg \Gamma_{0}$, where $\Gamma_{0}$ is the bare
level width. For interacting electrons the validity of this
approach (perturbation theory on $\Gamma_{0}$) for high-T regime
of electron transport was proved for $g<1/2$ (strong interaction)
\cite{15} and when $1-g\ll 1$ (weak interaction) \cite{16}.

We found that at low temperatures: $\Gamma_{0}\ll T\ll \hbar
w_{0}$ ($\hbar w_{0}$ is the characteristic energy of vibrons) the
peak conductance scales with temperature accordingly to Furusaki
prediction \cite{14}: $G(T)\propto (\Gamma_{\lambda}/T)
(T/\Lambda)^{1/g-1}$ ($\Lambda\simeq \varepsilon_{F}$ is the
Luttinger liquid cutoff energy). The influence of electron-phonon
interaction in low-T region results in renormalization of bare
level width: $\Gamma_{\lambda}= \Gamma_{0}\exp(-\lambda^{2})$,
where $\lambda$ is the dimensionless constant of electron-phonon
interaction. In the intermediate temperature region:
 $\hbar w_{0}\leq T\leq \lambda^{2}\hbar w_{0}$, ($\lambda\gg 1$),
Furusaki scaling is changed to $G(T)\propto (T)^{1/g-3/2}$ and at
high temperatures when all inelastic channels for electron
tunnelling are open we again recovered Furusaki scaling with
nonrenormalized level width ($\Gamma_{0}$).

For nonlinear regime of electron tunnelling we showed that
zero-bias peak in differential conductance, presenting elastic
tunnelling, is suppressed by Coulomb correlations in the leads.
This is manifestation of the Kane-Fisher effect \cite{14,15}. When
interaction is moderately strong ($1/2\leq g<1$) the dependence of
differential conductance on bias voltage is non-monotonous due to
the presence of satellite peaks. For $g>1/2$ the zero-bias peak
can be even more suppressed than the satellite peaks, which
dominate in this case. This is the manifestation of the interplay
between the Luttinger liquid effects in the leads and the
electron-phonon coupling in the dot . For strong interaction
$g<1/2$ satellites are also suppressed and the differential
conductance at low temperatures ($T\ll \hbar w_{0}$) scales as
$dI/dV\propto V^{1/g-2}$. This scaling coincides with the Furusaki
prediction, where temperature is replaced by the driving voltage
($eV$) which becomes the relevant energy scale for $eV\gg T,\hbar
w_{0},\Gamma$. It means that the influence of vibrons on the
resonant electron tunnelling through a vibrating quantum dot can
be observed only for weak or medium strong interaction ($1/2<g<1$)
in the leads.

\section{\bf The Model}

The Hamiltonian of our system (vibrating quantum dot weakly
coupled to Luttinger liquid leads, (see Fig.1) consists of three
parts
\begin{equation}
{\cal H}={\cal H}_{LL}+{\cal H}_{QD}+{\cal H}_{T} \ .
\label{1}
\end{equation}
Here ${\cal H}_{LL}=\sum_{j=L,R}{\cal H}_{l}^{(j)}$ describes
quantum wires adiabatically connected to electron reservoirs.
Quantum wires (left-L and right-R) are supposed equal and modelled
by Luttinger liquid Hamiltonians with equal Luttinger liquid
parameters $1/g_{L(R)}$: $1/g_{L}=1/g_{R}=1/g$ (see e.g.\cite{14})
\begin{equation}
{\cal H}_{l}^{L(R)}={\cal H}_{l}=\hbar v_{c}
\int_{0}^{\infty}a_{k}^{+}a_{k}kdk \ .
\label{2}
\end{equation}
Here $a_{k}^{+}$($a_{k}$) are the creation (annihilation)
operators of bosons which describe the charge density fluctuations
propagating in the leads with velocity $v_{c}\sim v_{F}$. These
operators satisfy canonical bosonic commutation relations
$[a_{k},a_{k'}^{+}]=\delta (k-k')$. In what follows we consider
for simplicity the case of spinless electrons.

The Hamiltonian of vibrating single level quantum dot takes the
form (see e.g.\cite{10})
\begin{equation}
{\cal H}_{QD}=\varepsilon_{0}
f^{+}f+\varepsilon_{i}(b^{+}+b)f^{+}f+\hbar w_{0}b^{+}b \ ,
\label{3}
\end{equation}
where $\varepsilon_{0}$ is the energy of electron level on the
dot, $\hbar w_{0}$ is the energy of vibrons, $\varepsilon_{i}$ is
the electron-vibron interaction energy, $f^{+}$($f$) and
$b^{+}$($b$) are fermionic ($f$) and bosonic ($b$) creation
(annihilation) operators with canonical commutation relations
$\{f,f^{+}\}=1$, $[b,b^{+}]=1$.

The tunnelling Hamiltonian is given by standard expression
\begin{equation}
{\cal H}_{T}=\sum_{j=L,R}\{t_{j}f^{+} \Psi(j)+h.c.\} \ ,
\label{4}
\end{equation}
where $t_{j}$ is the electron tunnelling amplitude and $\Psi(j)$,
$j=L,R$ is the annihilation  operator of electron at the end point
of L(R)-electrode. This operator could be written in a "bosonised"
form (according to \cite{14})
\begin{equation}
\Psi(L(R))=\sqrt{\frac{2}{\pi \alpha}}\cdot
\exp\left[\int_{0}^{\infty}dk \frac{e^{-\alpha
k/2}}{\sqrt{2K_{\rho} k}} \cdot(a_{k}-a_{k}^{+})\right] \ ,
\label{5}
\end{equation}
Here $\alpha$ is a short-distance cutoff of the order of the
reciprocal of the Fermi wave number $k_{F}$ and
$K_{\rho}=(2/g-1)^{-1}$ is the interaction parameter in the
"fermionic" form of the Luttinger liquid  Hamiltonian (2), it
defines the Luttinger liquid parameter $g$ which is varied between
$0$ and $1$: the case $g=1$ describes the "noninteracting"
(Fermi-liquid) leads, than in the case $g\rightarrow 1$ the
interaction in the leads goes to infinity.

\begin{figure}
\includegraphics[height=8 cm,width=8.6 cm]{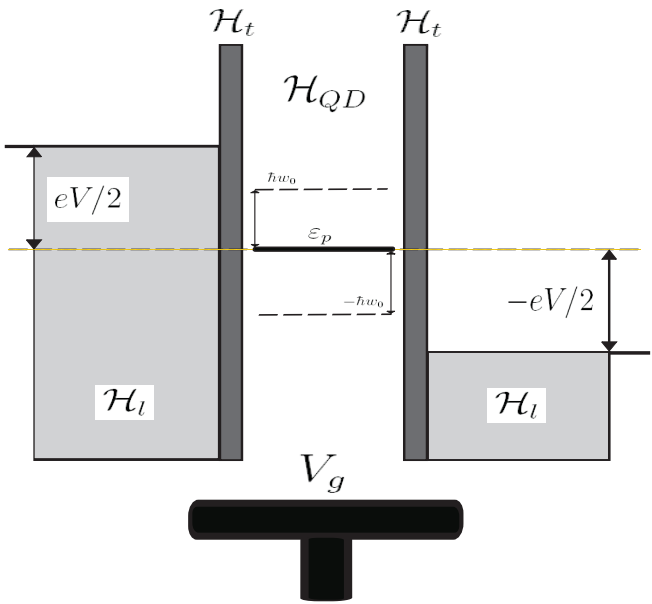}
\caption{The schematic picture of the two-terminal electron transport
through a vibrating quantum dot weakly coupled (via narrow
dielectric regions $H_t$) to the Luttinger liquid leads (${\cal
H}_{l}^{L(R)}={\cal H}_{l}$) with the chemical potentials
$\varepsilon_F\pm eV/2$ ($V$ is the driving voltage). All the
energies are counted from the Fermi energy, which chosen to be
zero. Electrons tunnel from one lead to another by hopping on and
off the dot level with the energy $\varepsilon_P$ (elastic
channel) and due to electron-vibron coupling they can emit or
absorb vibrons (vibron-assisted tunnelling). Inelastic channels
are represented as side-levels with energies
$\varepsilon_{P\pm}=\varepsilon_P\pm\hbar\omega_0$. The position
of the dot levels with respect to the Fermi energy can be
uniformly shifted by applying voltage $V_g$ to the "gate"
electrode.}
\end{figure}

Hamiltonian (3) is "diagonalized" to ${\cal
H}_{d}=\varepsilon_{P}f^{+}f+\hbar w_{0}b^{+}b$ by the unitary
transformation (see e.g. \cite{17}) $U=\exp(i\lambda pn_{f})$,
where $p=i(b^{+}-b)/\sqrt{2}$, $n_{f}=f^{+}f$ and the
dimensionless parameter $\lambda=-\sqrt{2}\varepsilon_{i}/\hbar
w_{0}$ characterizes electron-vibron coupling. The unitary
transformation results in: (i) the shift of fermionic level
(polaronic shift)
$\varepsilon_{P}=\varepsilon_{0}-\varepsilon_{i}^{2}/\hbar w_{0}$
and, (ii) the replacement of tunnelling amplitude in (3)
$t_{j}\Rightarrow t_{j}\cdot \exp(-i\lambda \cdot p)$. The model
Eqs.(\ref{1})-(\ref{5})can not be solved exactly and one needs to
exploit certain approximations to go further.

We will use "master equation" approximation (see e.g. \cite{5}) to
evaluate the average current and noise power in our model. It is
in this approximation that average current separately for the
model with interacting leads \cite{14} and for vibrating quantum
dot with noninteracting leads \cite{18} was calculated earlier.
Master (rate) equation approach exploits such quantities as the
probability for electron to occupy dot level and the transition
rates. It neglects quantum interference in electron tunnelling and
therefore describes only the regime of sequential electron
tunnelling which is valid when the width of electron level
$\Gamma_{0}\ll \min(T,eV)$ . In other words, in our case "master
equation" approach is equivalent to the lowest order of
perturbation theory in $\Gamma_{0}$.

For interacting electrons the validity of master equation approach
for high-T regime of resonant electron tunnelling can be justified
for strong repulsive interaction $g<1/2$ \cite{14}. It is correct
also for weak interaction $1-g\ll 1$ as one can check by comparing
the results of Ref.\cite{14} and Ref. \cite{16}, where resonant
tunnelling through a double-barrier Luttinger liquid was
considered for weak electron-electron interaction. Notice, that
the results \cite{18} of exact solution known for $g=1/2$, where a
mapping to free-fermion theory can be used \cite{5}, do not agree
with the high-T scaling of $G(T)$ \cite{14} extrapolated to this
special point $g=1/2$. The free-fermion scaling $G(T)\propto
T^{-1}$ found for $g=1/2$ (master equation approach predicts
$T$-independent value \cite{14}) could be a special feature of
this exactly solvable case. We will assume that beyond the close
vicinity to $g=1/2$ the master equation approach for high-T
behavior of conductance is a reasonable approximation.

\section{\bf Transition rates and the average current}

In master equation approach the average current through a single
level quantum dot expressed in terms of transition rates takes the
form
\begin{equation}
I=e
\frac{\Gamma_{01}^{R}\Gamma_{10}^{L}-\Gamma_{01}^{L}\Gamma_{10}^{R}}
{\Gamma_{\Sigma}} \ ,
\label{6}
\end{equation}
where $\Gamma_{01}^{R(L)}$ is the rate of electron tunnelling from
the dot to right (left) electrode, $\Gamma_{10}^{R(L)}$ describes
the reverse process and $\Gamma_{\Sigma}=\Gamma_{01}+\Gamma_{10}$
, $\Gamma_{if}=\Gamma_{if}^{L}+\Gamma_{if}^{R}$ $(i,f=0,1)$. To
evaluate these rates in our approach we will use Fermi "Golden
Rule" (quantum mechanical perturbation theory) for tunnelling
Hamiltonian obtained from Eq.(4) after the unitary transformation:
${\cal H}_{T}\Rightarrow {\cal H}_{t}$
\begin{equation}
{\cal H}_{t}=\sum_{j=L,R}\{t_{j}\Psi^{+}(j)f\exp(-i\lambda \cdot
p)+h.c.\} \ . \label{7}
\end{equation}
The standard calculation procedure results in the following
expressions for tunnelling rates
\begin{eqnarray}
\Gamma_{10}^{(j)}=\left|\frac{t_{j}}{\hbar}\right|^{2}
\int_{-\infty}^{\infty}dt\langle V(t)V^{+}(0)\rangle_{b}\langle
\Psi_{j}^{+}(t)\Psi_{j}(0)\rangle_{f}\nonumber\\
\cdot \exp(i(\varepsilon_{P}-\varepsilon_{F}+eV_{j})t/\hbar) \ ,
\label{8}
\end{eqnarray}
\begin{eqnarray}
\Gamma_{01}^{(j)}=\left|\frac{t_{j}}{\hbar}\right|^{2}
\int_{-\infty}^{\infty}dt\langle V^{+}(t)V(0)\rangle_{b}\langle
\Psi_{j}(t)\Psi_{j}^{+}(0)\rangle_{f}\nonumber\\
\cdot \exp(-i(\varepsilon_{P}-\varepsilon_{F}+eV_{j})t/\hbar) \ ,
\label{9}
\end{eqnarray}
where $V_{L}-V_{R}=V$ is the bias voltage and $j=L,R$. Notice that
in the perturbation calculation on the bare level width
$\Gamma_{0}\propto|t_{L,R}|^{2}$, we neglect the level width in
the Green function of the dot level. Besides, in this
approximation averages over bosonic and fermionic operators in
formulas for tunnelling rates are factorized and, thus, the
averages $\langle\ldots\rangle_{b}$ over bosonic variables
\begin{equation}
V=\exp(-i\lambda \cdot p) \ , \ \ \ p=\frac{i}{\sqrt{2}}[b^{+}-b]
\label{10}
\end{equation}
can be calculated with the quadratic Hamiltonian ${\cal
H}_{b}=\hbar w_{0}b^{+}b$. In what follows we will assume that
vibrons are characterized by equilibrium distribution function:
$n_{b}(T)=[\exp(\hbar w_{0}/T)-1]^{-1}$. Averages
$\langle\ldots\rangle_{f}$ over fermionic operators in Eqs.(8),(9)
are calculated with the Luttinger liquid Hamiltonian (2). The
corresponding correlation functions in Eqs.(8),(9) are well known
in the literature (see e.g.\cite{10,14})
\begin{eqnarray}\nonumber\\
\langle V(t)V^{+}(0)\rangle_{b}=\exp(-\lambda^{2}(1+2n_{B}))\nonumber\\
\cdot\left\{\sum_{l=-\infty}^{\infty}I_{l}\left[2\lambda^{2}\sqrt{n_{B}(1+n_{B})}\right]\exp[-ilw_{0}(t+i\hbar/2T)]\right\}
\nonumber\\
\end{eqnarray}

\begin{equation}
\langle \Psi_{j}^{+}(t)\Psi_{j}(0)\rangle_{f}\simeq
\frac{\Lambda}{\hbar v_{F}}\cdot \left\{\frac{i\Lambda}{\pi
T}\cdot \sinh \left[\frac{\pi T\cdot t}{\hbar}
\right]\right\}^{-1/g} \ .
\label{12}
\end{equation}
Here $I_{l}(z)$ is a modified Bessel function, $\Lambda\sim
\varepsilon_{F}$ is a ultraviolet cutoff energy, $g$ is the
Luttinger liquid correlation parameter.

By putting correlation functions (11),(12) in Eqs.(8),(9) and
evaluating time integrals we get the following equations for
tunnelling rates $\Gamma^{(j)}_{+}=\Gamma^{(j)}_{10}$ and
$\Gamma^{(j)}_{-}=\Gamma^{(j)}_{01}$, ($j=L,R$)

\begin{eqnarray}
\nonumber\\
\Gamma^{(j)}_{\pm}=\frac{\Gamma_{j}}{2\pi \hbar}\frac{\exp\left[-\lambda^{2}\coth(\hbar w_{0}/2T)\pm
\Delta_{j}/2T\right]}{\Gamma(1/g)}
\nonumber\\
\cdot\left(\frac{2\pi T}{\Lambda}\right)^{1/g-1}\sum_{l=-\infty}^{\infty}
I_{l}\left[\frac{\lambda^{2}}{\sinh\left(\hbar
w_{0}/2T\right)}\right]
\nonumber\\
\cdot\left|\Gamma\left(\frac{1}{2g}+i\frac{(\pm\Delta_{j}-\hbar
w_{0}l)}{2\pi T}\right)\right|^{2}
\nonumber\\
\end{eqnarray}

where $\Gamma_{L(R)}$ is the partial level width (see, for
example,\cite{14} )
 $\Gamma_{j}=\left(2\pi c_{j}t_{j}^{2}/\alpha \Lambda\right)\langle f^{+}f \rangle=const_{j}$
 ($j=L,R$), $\Gamma_{L}+\Gamma_{R}=\Gamma_{0}$,
 $\Delta_{j}=\varepsilon_{F}-\varepsilon_{P}+eV_{j}$;
here $\Gamma(z)$ is Gamma function.

At first we consider different limiting cases when it is possible
to obtain simple analytical expressions for the average current
(6). Notice, that electric current depends on the gate voltage
$V_{g}$ through the corresponding dependence of level energy
$\varepsilon_{P}(V_{g})$. It is convenient for the further
analysis to choose the value of gate voltage at which the current
at low bias is maximum as:
$\varepsilon_{P}(V_{g})=\varepsilon_{F}$. In what follows we also
put $V_{L}=-V_{R}=V/2$.

For noninteracting leads $(g=1)$ and noninteracting quantum dot
$(\lambda=1)$ it is easy to derive from Eqs.(6),(13) the
well-known formula for the maximum (resonant) current at
temperatures $T\gg \Gamma_{L(R)}$
\begin{equation}
I(V)\simeq \frac{e\Gamma}{\hbar}\tanh\left(\frac{eV}{4T}\right)\ ,
\label{14}
\end{equation}
where $\Gamma=\Gamma_{L}\Gamma_{R}/(\Gamma_{L}+\Gamma_{R})$ is the
effective level width. It is evident that at high voltages: $eV\gg
T$ the current through a single level dot does not depend on the
bias voltage and its value is totally determined by the effective
level width $\Gamma$.

For a vibrating quantum dot $(\lambda\neq 0)$ weakly coupled to
noninteracting leads $(g=1)$ our approach reproduces the results
of Ref.\cite{10}. In the temperature region we are interesting in
$(T\gg \Gamma_{L(R)})$ the general formulae derived in \cite{10}
can be rewritten in a more clear and compact form. In particular,
by using for $g=1$ in Eq.(11) the well-known representation for
Gamma function (see e.g.\cite{19})
\begin{equation}
\left|\Gamma\left(\frac{1}{2}+iz\right)\right|^{2}=\frac{\pi}{\cosh(\pi
z)} \ ,
\label{15}
\end{equation}
it is easy to obtain the following expression for the maximum
(peak) conductance
\begin{equation}
G_{\lambda}(T)=G(T)\cdot F_{\lambda}\left(\frac{\hbar
w_{0}}{T}\right) \ ,
\label{16}
\end{equation}
where
\begin{equation}
G(T)\simeq \frac{\pi}{2}G_{0}\frac{\Gamma}{T}\ \ \ , \ \ \
G_{0}=\frac{e^{2}}{h}
\label{17}
\end{equation}
is the standard resonance conductance of a single-level quantum
dot at $T\gg \Gamma_{L(R)}$. The dimensionless function
$F_{\lambda}(x)$ takes the form
\begin{eqnarray}
F_{\lambda}(x)=\exp[-\lambda^{2}(1+n_{B}(x))]\nonumber\\
\cdot\left\{I_{0}(z(x))+2\cdot \sum_{l=1}^{\infty}
\frac{I_{l}(z(x))}{\cosh^{2}(lx/2)}\right\} \ ,
\label{18}
\end{eqnarray}
here $z(x)\equiv 2\lambda^{2}\sqrt{n_{B}(x)[1+n_{B}(x)]}$ and
$n_{B}(x)=(\exp(x)-1)^{-1}$. At low temperature region $\Gamma\ll
T\ll\hbar w_{0}$, when there are no thermally activated vibrons in
the dot $(n_{B}\ll1)$ only the first term in the brackets
contribute to the sum and: $F_{\lambda}(T\ll\hbar w_{0})\simeq
\exp(-\lambda^{2})$. We see, that zero-point fluctuations of the
dot position result in renormalization  of the level width
$\Gamma_{\lambda}=\Gamma\cdot \exp(-\lambda^{2})$. For strong
electron-vibron coupling this phenomenon (polaronic narrowing of
level width) leads to polaronic  (Franck-Condon) blockade of
electron transport through vibrating quantum dot \cite{11}. The
temperature behavior of peak conductance (16) was considered in
Ref.\cite{20}.

Now we will study the general case when interacting quantum dot
$(\lambda\neq 0)$ is connected to interacting leads $(g<1)$.
Analytical expressions for conductance in this case can be
obtained in the limits of low $(\Gamma_{L(R)}\ll T\ll \hbar
w_{0})$ and high $(T\gg \hbar w_{0})$ temperatures.

At low temperatures the main contribution to the sum over $"l"$ in
Eq.(13) comes from elastic transition $l=0$. All inelastic
channels $(l\neq 0)$ are exponentially suppressed for $eV,T\ll
\hbar w_{0}$. At $T\ll g\hbar w_{0}$ the peak conductance takes
the form
\begin{equation}
G(T)\simeq \frac{\sqrt{\pi}}{2}G_{0}\frac{\Gamma_{\lambda}}{T}
\left[\frac{\Gamma\left(1/2g\right)}{\Gamma\left(1/2g+1/2\right)}\right]
\left(\frac{\pi T}{\Lambda}\right)^{1/g-1} \ .
\label{19}
\end{equation}
We see that at low temperatures conductance scales with
temperature according to Furusaki's prediction \cite{14}:
$G(T)\sim T^{1/g-2}$. The influence of electron-vibron coupling
results in multiplicative renormalization of bare level width
$\Gamma_{\lambda}=\Gamma\exp(-\lambda^{2})$.

At high temperatures: $T\gg \hbar w_{0}$ one can use the well
known asymptotic expansion for Bessel function $I_{l}(z)\simeq
\exp(z)/\sqrt{2\pi z}$, which can be used in summation
Eqs.(\ref{18}), (13) until $l^{2}\leq z$. Besides, in this
temperature region  the summation in Eq.(13), can be replaced by
integration and the corresponding integral can be taken exactly
\begin{equation}
\int_{-\infty}^{\infty} |\Gamma(a+iz)|^{2}dz=\pi 2^{1-2a}
\Gamma(2a) \ .
\label{20}
\end{equation}
This allows us to derive the following expression for the
temperature dependence of peak conductance in the intermediate
temperature region $\hbar w_{0}\ll T \leq \lambda^{2}\hbar w_{0}$,
$(\lambda\geq 1)$
\begin{equation}
G(T)\simeq \frac{\pi}{2}G_{0}
\left[\frac{\Gamma\exp\left(-\lambda^{2}\hbar
w_{0}/4T\right)}{\lambda \sqrt{\hbar w_{0}T}}\right]
\left(\frac{\pi T}{\Lambda}\right)^{1/g-1} \ .
\label{21}
\end{equation}
Notice that in the considered temperature region the  polaronic
blockade is already partially lifted $
\Gamma_{\lambda}(T)=\Gamma\exp\left(-\lambda^{2}\hbar
w_{0}/4T\right)\sim \Gamma$ at $T\sim \lambda^{2}\hbar w_{0}$ and
conductance scales with temperature as $G(T)\sim T^{1/g-3/2}$. At
last, at temperatures $T\gg \lambda^{2}\hbar w_{0}$ when all
inelastic channels for electron transport are open, the polaronic
blockade is totally lifted \cite{20} and we reproduce again
Furusaki scaling. It is clear from our asymptotic formulae
(19),(21) that both in low- and in high- temperature regions the
contributions of electron-electron and electron-vibron
interactions to the conductance are factorized. In general case
these contributions are not factorized, as one can see from
Eqs.(8),(9) and from Eq.(13) for tunnelling rates, and we can
expect interplay of Kane-Fisher effect and the effect of
phonon(vibron)-assisted tunnelling.

To see this interplay we consider nonlinear (differential)
conductance $G(V)=dI/dV$. It is well known that Kane-Fisher effect
is pronounced for the energies close to the Fermi energy. For
differential conductance it means that the zero-bias (elastic)
resonance peak is suppressed with the increase of
electron-electron interaction, while satellite peaks are less
affected by the interaction. When electron-electron interaction is
weak or moderately strong $(1/2\leq g<1)$ the dependence of
differential conductance on the bias voltage (for $\lambda\simeq
1$) is not a monotonous function due to the presence of satellite
peaks (see Figs.2,3). 

\begin{figure}
\includegraphics[height=6 cm,width=8.6 cm]{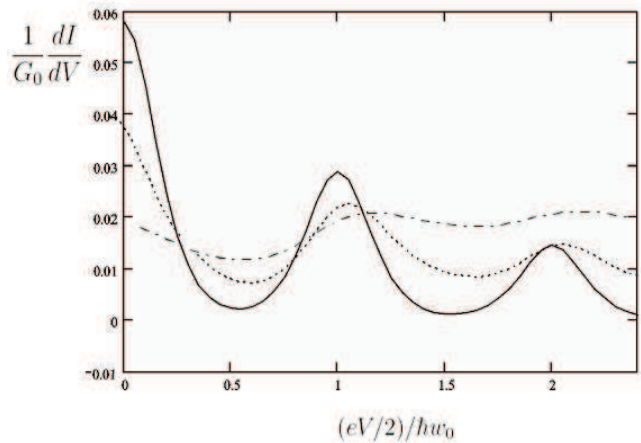}
\caption{Differential conductance (in the units of $G_{0}$) as a function
of driving voltage (in the units of $\hbar w_{0}$) for the case $g>1/2$. Here we put
$\Gamma/kT=0.01$; $\hbar w_{0}/kT=10$; $\lambda^{2}=1$; and tune
the level energy to the resonant position $\varepsilon_{0}=\lambda^{2}\hbar w_{0}$ ($\varepsilon_{P}=0$).
Solid line corresponds to the case of noninteracting
leads: $g=1$. Dot line corresponds to value $g=0.8$, while dash-dot line - to $g=0.6$. 
Figure shows how zero-bias (elastic) resonance peak is gradually
suppressed with the increase of electron-electron correlations
(decrease of Luttinger liquid parameter $g$) while the satellite
peaks survive until $g>1/2$.}
\end{figure}

\begin{figure}
\includegraphics[height=6 cm,width=8.6 cm]{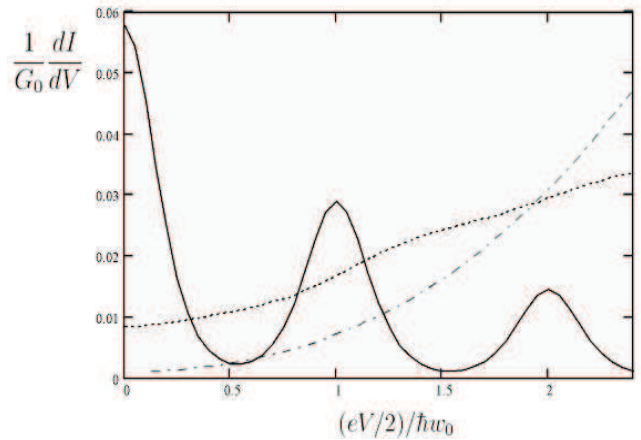}
\caption{Differential conductance (in the units of $G_{0}$) as a function
of driving voltage (in the units of $\hbar w_{0}$) for the case $g<1/2$. Here $\Gamma/kT=0.01$; $\hbar w_{0}/kT$; $\lambda^{2}$; and  $\varepsilon_{0}$ are the same as on Fig.2.
Solid line corresponds to the case of noninteracting
leads: $g=1$; dot line corresponds to value $g=0.45$; dash-dot line - to $g=0.25$.
One can see from the figure that for $g<1/2$ the resonance-like behavior of differential conductance disappears and
conductance scales as a power-law of the bias voltage.}
\end{figure}

The resonant behavior disappears for strong
interaction $g<1/2$ (Figs.2,3), when at low temperatures $T\ll
\hbar w_{0}$ differential conductance scales with bias voltage as
$G(V)\propto V^{1/g-2}$ in accordance with the Luttinger liquid
prediction for nonlinear electron transport through a single-level
quantum dot. For instance, if we put $1/g=n,\, (n=2,3,\ldots)$ and
tune the level energy to the resonance point
$\varepsilon_{0}=\lambda^{2}\hbar w_{0}$ -("resonant" location of
the level in the presence of "polaronic" shift), we obtain for the
differential conductance $G(V)$ the following expression for
$eV/\hbar w_{0}\gg (1/g-2)$
\begin{equation}
G(V)\simeq 4\pi G_{0}\frac{\Gamma_{\lambda}}{\Lambda}\cdot
\left[\frac{1}{(n-1)!}\right]\cdot
\left(\frac{eV}{2\Lambda}\right)^{n-2} \ ,
\label{22}
\end{equation}
where $\Gamma_{\lambda}=\Gamma\exp(-\lambda^{2})$. One can readily
see that expression (22) reproduces Furusaki temperature scaling
Eq.(19) when $eV$ is replaced by $T$.

Analogous interplay of Kane-Fisher and polaronic effects one can
see on Figs.4,5, where differential conductance is plotted as a
function of level energy $\varepsilon_0$ (or, equivalently, as a
function of gate voltage). For noninteracting leads $(g=1)$ the
resonance conductance peaks correspond to the level positions at
$\varepsilon_{P\pm}=\varepsilon_{F}\pm eV/2$ (in our plot we put:
$eV=5\hbar w_{0}$). This elastic resonance peak is suppressed by
electron-electron interaction in the leads $(g<1)$. The dependence
$G(V_{g})$ for weak and moderately strong interaction still
reveals resonance structure with 4 satellites in our case (see Fig.4).
The inelastic resonance peaks disappear at $g<1/2$ and maximum of
differential conductance corresponds at $g\ll 1$ to the level
position at $\varepsilon_{P}(V_{g})=\varepsilon_{F}$, that is
exactly in the middle between chemical potentials of left and
right electrodes (see Fig.5).

\begin{figure}
\includegraphics[height=6 cm,width=8.6 cm]{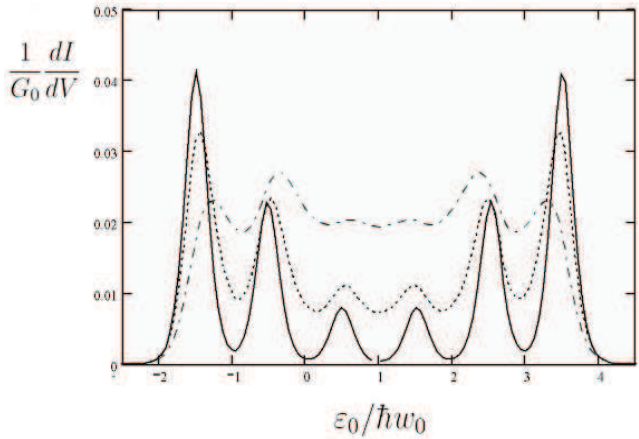}
\caption{Differential conductance (in the units of $G_{0}$) as a function
of level energy $\varepsilon_0$, counted from the Fermi energy for the case $g>1/2$.
Solid line corresponds to the case of noninteracting
leads: $g=1$; dot line corresponds to value $g=0.8$; dash-dot line - to $g=0.6$. 
The bias voltage $eV/\hbar\omega_0=5$ is sufficiently high to
excite vibrons and to support electron transport through inelastic
channels. All parameters are the same as for Fig.2.}
\end{figure}

\begin{figure}
\includegraphics[height=6 cm,width=8.6 cm]{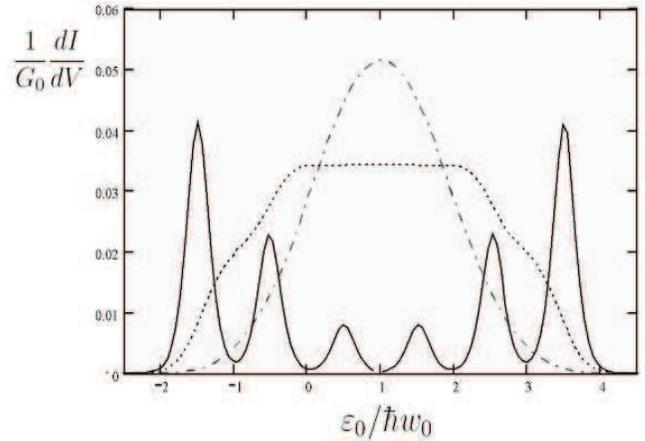}
\caption{Differential conductance (in the units of $G_{0}$) as a function
of level energy $\varepsilon_0$, counted from the Fermi energy for the case $g<1/2$.
Solid line corresponds to the case of noninteracting leads: $g=1$; dot line corresponds to value $g=0.45$; dash-dot line - to $g=0.25$. The bias voltage $eV/\hbar\omega_0=5$ is sufficiently high to
excite vibrons and to support electron transport through inelastic
channels. All parameters are the same as for Fig.3.}
\end{figure}

It is important to stress here once more that for moderately
strong electron-electron interaction in the leads the inelastic
tunnelling can dominate in electron transport. One can see from
Figs.2-5 that there is region of coupling constants when the first
satellite peak is higher than the "main" (zero-bias) resonant
peak, which corresponds to elastic ($l=0$) tunnelling channel. It
is the most significant prediction, we have made in this paper.

\section{\bf The noise power}

The knowledge of tunnelling rates Eqs.(8,9) allows us to evaluate
not only the average current Eq.(6) but the noise power as well.
We will follow the method developed in Refs. \cite{21,22} where
quantum noise was calculated for resonant electron transport
through a quantum dot weakly coupled to noninteracting electrodes.

The noise power is defined (see e.g. \cite{23}) as
\begin{equation}
S(w)=2\int_{-\infty}^{\infty} dt\exp(iwt) \langle\Delta I(t)\Delta
I(0)\rangle \ ,
\label{23}
\end{equation}
where $\Delta I(t)=I(t)-I$ ($I$ is the average current). The noise
defined in Eq.(23), in the case of sequential tunnelling through a
quantum dot, can be expressed in terms of tunnelling rates. For a
single level quantum dot this formula for low frequency noise
$S=S(w=0)$ takes the form
\begin{eqnarray}
S=2eI-\frac{4I^{2}}{\Gamma_{\Sigma}}+4e^{2}
\frac{\Gamma_{01}^{L}\Gamma_{10}^{R}}{\Gamma_{\Sigma}} \ ,
\label{24}
\end{eqnarray}
here the average current $I$ is determined by Eq.(6). The noise
power Eq.(24) depends on temperature and bias voltage $S(T,V)$ and
contains both thermal (Jonson-Nyquist) noise $S_{JN}(T)\equiv
S(T,V=0)=4TG(T)$ ($G$ is the conductance) and the non-equilibrium
(shot) noise $S_{sh}(T,V)$. Since the thermal noise is totally
determined by temperature dependence of conductance, we will study
in what follows only shot noise and Fano factor $F=S_{sh}/2eI$. In
particular, Fano factor in our case can be represented as follows
\begin{equation}
F=\left\{1-\frac{2I}{e\Gamma_{\Sigma}}+\frac{2e}{I}\left(\frac{\Gamma_{01}^{L}
\Gamma_{10}^{R}}{\Gamma_{\Sigma}}-\frac{TG}{e^{2}}\right)\right\}
 \ .
\label{25}
\end{equation}
For noninteracting leads $(g=1)$ and noninteracting quantum dot
$(\lambda=0)$ one readily gets from Eqs.(13),(24) a simple
expression for the "full" noise $(S)$ of a single electron
transistor (SET). On resonance
$\varepsilon_{P}(V_{g})=\varepsilon_{F}$ and at temperatures $T\gg
\Gamma$ one finds
\begin{eqnarray}
S=\frac{2e^{2}\Gamma}{\hbar}\tanh\left(\frac{eV}{4T}\right)
\left[1-\frac{2\Gamma}{\Gamma_{\Sigma}}\tanh\left(\frac{eV}{4T}\right)
\right]\nonumber\\
+\frac{e^{2}\Gamma}{\hbar}\left[\frac{\exp(-eV/4T)}{\cosh^{2}(eV/4T)}
\right]\ .
\label{26}
\end{eqnarray}

From Eq.(26) in the limit $V\rightarrow 0$ we obtain $S=
S_{JN}(T)$, where $S_{JN}(T)=e^{2}\Gamma/\hbar=4TG$ is the thermal
noise. In the opposite case $eV\gg T$ we rederive the well-known
formulae for the shot-noise and the Fano factor of a single level
quantum dot \cite{21,22,23}
\begin{equation}
S_{sh}=\frac{2e^2\Gamma}{\hbar}\left(1-\frac{2\Gamma}{\Gamma_{\Sigma}}\right)\,\,,\,\,
F=1-\frac{2\Gamma}{\Gamma_{\Sigma}}\ .
\label{27}
\end{equation}

These formulae (26),(27) can be also re-derived from the general
expression for the full noise of noninteracting electrons (see
e.g., Eq.(61) in Ref.\cite{23})

\begin{eqnarray}
S(V,T)=\frac{e^{2}}{\hbar}\int d\varepsilon T_{t}(\varepsilon)[f_{L}(1-f_{L})+f_{R}(1-f_{R})]\nonumber\\
+\frac{e^{2}}{\hbar}\int d\varepsilon T_{t}(\varepsilon)\left[1-T_{t}(\varepsilon)\right](f_{L}-f_{R})^{2}\ ,
\label{28}
\end{eqnarray}

where $T_{t}(\varepsilon)$ is the transmission coefficient and
$f_{j}=\left\{\exp\left[(\varepsilon-\mu_{j})/T\right]+1\right\}^{-1}$
is the equilibrium distribution function of electrons in the leads
($\mu_{j}$ is the chemical potential; $j=L,R$). In the case of
single level quantum dot $T_{t}(\varepsilon)$  takes the form
Breit-Wigner tunnelling probability
\begin{equation}
T_{t}(\varepsilon)=
\frac{\Gamma_{L}\Gamma_{R}}{\left(\varepsilon-\varepsilon_{P}
\right)^{2}+(\Gamma_{L}+\Gamma_{R})^{2}/4}\,
 \label{29}
\end{equation}
where $\Gamma\equiv
\Gamma_{L}\Gamma_{R}/\left(\Gamma_{L}+\Gamma_{R}\right)$. For a
weak tunnelling when $\Gamma_{L(R)}$ are the smallest energy
scales in the problem the Lorentzian shape of the Breit-Wigner
resonance shrinks to $\delta$-function
\begin{equation}
T_{t}(\varepsilon)|_{\Gamma_{L,R}\rightarrow 0}\simeq 2\pi
\Gamma\delta(\varepsilon-\varepsilon_{P}) \ .
 \label{30}
 \end{equation}
With the help of Eqs.(28),(30) for the resonance condition
$\varepsilon_{P}(V_{g})=\varepsilon_{F}$ we easily re-derive
Eqs.(26). (Notice, that in sequential tunnelling approach the
tunnelling transitions through the left and right barriers are
assumed to be weak and uncorrelated. Therefore we can safely
neglect $T_t^2$-term in Eq.(28)). It is evident from Eqs.(25),(27)
that for noninteracting electrons the Fano factor is
sub-Poissonian $(F\leq 1)$ and $F$ approaches $1$ for strongly
asymmetric junction $\Gamma_{L(R)}\gg \Gamma_{R(L)}$ and for
$eV\gg T$.

The master equation approach we have used in our analysis holds
when electron tunnelling amplitudes are small. For noninteracting
electrons this assumption is satisfied when electron energies are
far from the resonant energy level, i.e. $T_t(\varepsilon)\ll 1$.
The differential shot noise in this case as a function of bias
voltage or gate voltage behaves similarly to the differential
conductance. Notice however that due to different dependence on
temperature the shot noise unlike the thermal one even in
sequential tunnelling regime ($T\gg \Gamma$) can not be expressed
in terms of conductance.

\begin{figure}
\includegraphics[height=6 cm,width=8.6 cm]{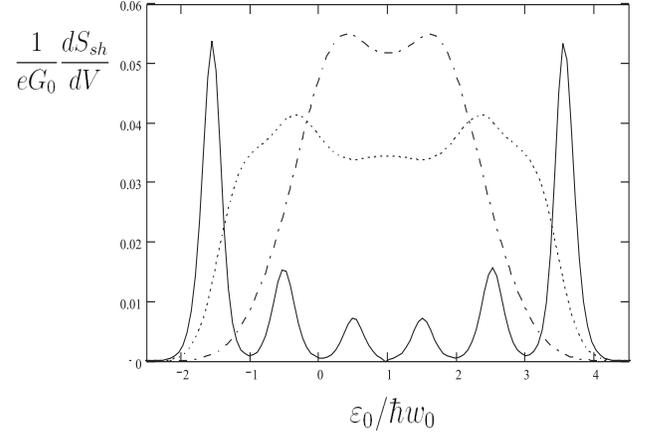}
\caption{Differential shot noise power (in the units of $eG_{0}$) as function of the level energy
$\varepsilon_0$ in the nonlinear transport regime (when $eV/\hbar\omega_0=5$) for the case $g<1/2$. 
Solid line corresponds to the case of noninteracting leads $g=1$; dot line corresponds to value $g=0.45$; dash-dot line - to $g=0.25$. Other parameters are the same as on Fig.5.}
\end{figure}

\begin{figure}
\includegraphics[height=6 cm,width=8.6 cm]{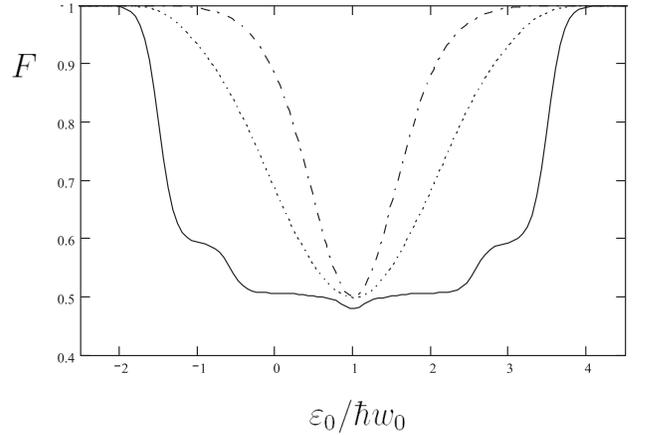}
\caption{Fano factor for the case $g<1/2$ as the function of the level energy
$\varepsilon_0$ in the nonlinear transport regime (when $eV/\hbar\omega_0=5$) which corresponds to Figs.5,6. 
Solid line corresponds to the case of noninteracting leads $g=1$; dot line corresponds to value $g=0.45$; dash-dot line - to $g=0.25$. All other parameters are the same as on Figs.5,6.}
\end{figure}

By comparing Fig.6 with Fig.5, one can see that the above
similarity is preserved for interacting electrons ($g\neq
1,\,\lambda\neq 0$) as well. The corresponding Fano factor which
is the "shot noise/current" ratio and thus is less sensitive to
the details of tunnelling process, for strong electron-electron
interaction exhibits a simple behavior (see Fig.7). It dips ($F
\simeq 1/2$) at symmetric (with respect to chemical potentials of
the leads) position of the dot level. Outside this region
$F\rightarrow 1$ (Poissonian noise). The width of the dip
decreases with the increase of interaction (see Fig.7).

\section{\bf Summary}

We considered the influence of interaction on transport properties
of molecular transistor which was modelled as a vibrating
single-level quantum dot weakly coupled to the Luttinger liquid
leads. We found interesting interplay between polaronic and
Luttinger liquid effects in our system. In particular it was shown
that for weak or moderately strong interaction ($1/2<g\leq 1$) the
differential conductance demonstrates resonance-like behavior and
for moderately strong interaction inelastic channels can even
dominate in electron transport through a vibrating quantum dot.
For strong interaction ($g\ll 1$) the resonant character of
vibron-assisted tunnelling disappears and the differential
conductance scales as a power law on temperature (linear regime
$T\gg eV$) or on bias voltage (nonlinear regime $eV\gg T$).

The authors would like to thank S.I.Kulinich for valuable
discussions. This work was partly supported by the joint grant of
the Ministries of Education and Science in Ukraine and Israel and
by the grant "Effects of electronic, magnetic and elastic
properties in strongly inhomogeneous nanostructures" of the
National Academy of Sciences of Ukraine.



\begin{thebibliography}{99}

\bibitem{1} A.Nitzan and M.A.Ratner, {\em Science} {\bf 300}, 1384 (2003).
\bibitem{2} M.Galperin, M.A.Ratner,  A.Nitzan, {\em J.Phys.:Cond.Matt.} {\bf 19}, 103201 (2007).
\bibitem{3} R.I.Shekhter, {\em Zh.Eksp.Teor.Fiz.} {\bf 68}, 623 (1975).
\bibitem{4} I.O.Kulik and R.I.Shekhter, {\em Zh.Eksp.Teor.Fiz.} {\bf 63}, 1400 (1972).
\bibitem{5} {\em Single Charge Tunneling}, edited by H.Grabert and M.H.Devoret,
            NATO ASI Ser.B, vol.294, Plenum Press, N.Y., (1992).
\bibitem{6} L.I.Glazman, R.I.Shekhter,{\em Zh.Eksp.Teor.Fiz.} {\bf 94}, 292 (1988)
            [{\em Sov.Phys.JETP} {\bf 67}, 163 (1988)].
\bibitem{7} S.Braig, K.Flensberg, {\em Phys.Rev.B} {\bf 68}, 205324 (2003).
\bibitem{8} L.Y.Gorelik, A.Isacsson, M.V.Voinova, B.Kasemo, R.I.Shekhter, M.Jonson,
            {\em Phys.Rev.Lett.} {\bf 80}, 4526 (1998).
\bibitem{9} R.I.Shekhter, L.Y.Gorelik, M.Jonson, Y.M.Galperin, V.M.Vinokur,
            {\em J.Comput.Theor.Nanosci.} {\bf 4}, 860 (2007).
\bibitem{10} U.Lundin, R.M.McKenzie, {\em Phys.Rev.B} {\bf 66}, 075303 (2002).
\bibitem{11} J.Koch, F. von Oppen, A.V.Andreev, {\em Phys.Rev.B}, {\bf 74}, 205438 (2006).
\bibitem{12} H.Park, J.Park, A.K.L.Lim, E.H.Anderson, A.P.Alivisatos, P.L.McEuen,
             {\em Nature} {\bf 407}, 57 (2000).
\bibitem{13} H.W.Ch.Postma,T.Teepen, Z.Yao, M.Grifoni, C.Dekker, {\em Science} {\bf 239}, 76 (2001).
\bibitem{14} A.Furusaki, {\em Phys.Rev.B} {\bf 57}, 7141 (1998).
\bibitem{15} C.L.Kane, M.P.A.Fisher, {\em Phys.Rev.B} {\bf 46}, 15233 (1992).
\bibitem{16} Yu.V.Nazarov and L.I.Glazman, {\em Phys.Rev.Lett.} {\bf 91}, 126804 (2003).
\bibitem{17} G.D.Mahan, {\em Many-Particle Physics}, Plenum Press, New York (1990).
\bibitem{18} A.Komnik and A.O.Gogolin, {\em Phys.Rev.Lett.} {\bf 90}, 246403 (2003).
\bibitem{19} "I.S.Gradshtein, and I.M.Ryzhik, {\em Tables of Integrals, Series and
             Products}, Academic Press, NY (1965).
\bibitem{20} I.V.Krive, R.Ferone, R.I.Shekhter, M.Jonson, P.Utko, J.Nygard, {\em New J.Phys.}, (2008) to be published.
\bibitem{21} S.Hershfield, J.H.Davies, P.Hyldgaard, C.J.Stanton, J.W.Wilkins, {\em Phys.Rev.B} {\bf 47}, 1967 (1993).
\bibitem{22} I.Djuric, B.Dong, H.L.Cui, {\em J.Appl.Phys.} {\bf 99}, 063710 (2006).
\bibitem{23} Y.M.Blanter, M.Buttiker, {\em Shot Noise in Mesoscopic Conductors}, {\em Phys.Rep.} {\bf 336}, pp.1-166, (2000).

\end{thebibliography}
\end{document}